\documentclass[twocolumn,showpacs,superscriptaddress,aps,prb]{revtex4}
\usepackage{graphicx}
\usepackage{longtable}
\begin{document}
%%%%%%%%%%%%%%%% T I T L E  &  A U T H O R S%%%%%%%%%%%%%
\title{Structural transition in Na$_x$CoO$_2$ with $x$ near
0.75 due to Na rearrangement}
\author{Q. Huang}
\affiliation{NIST Center for Neutron Research, NIST, Gaithersburg,
MD 20899-8562}
\author{B. Khaykovich}
\affiliation{Center for Materials Science and Engineering,
Massachusetts Institute of Technology, Cambridge, MA 02139}
\affiliation{Department of Physics, Massachusetts Institute of
Technology, Cambridge, MA 02139}
\author{F.C. Chou}
\affiliation{Center for Materials Science and Engineering,
Massachusetts Institute of Technology, Cambridge, MA 02139}
\author{J.H. Cho}
\affiliation{Department of Physics, Massachusetts Institute of
Technology, Cambridge, MA 02139}
\author{J. W. Lynn}
\affiliation{NIST Center for Neutron Research, NIST, Gaithersburg,
MD 20899-8562}
\author{Y.S. Lee}
\affiliation{Center for Materials Science and Engineering,
Massachusetts Institute of Technology, Cambridge, MA 02139}
\affiliation{Department of Physics, Massachusetts Institute of
Technology, Cambridge, MA 02139}

\date{\today}

\begin{abstract}
We report neutron powder diffraction measurements on a series of
Na$_x$CoO$_2$ samples with $x$ near 0.75 which were prepared under
different synthesis conditions.  Two different crystal structures
for the samples are observed at room temperature.  The two
structures differ primarily by a shift of a large fraction of the
Na ions from a high-symmetry position to a lower-symmetry
position.  Close inspection of the refinement parameters indicates
that the presence of either structure depends sensitively on the
Na content $x$, with $x\simeq 0.75$ as the critical concentration
separating the two phases. By raising the temperature to around
$T\simeq 323$~K, the high-symmetry structure can be converted to
the lower-symmetry structure.  The transition is reversible, but
there is significant hysteresis.  We discuss the effects of this
structural transition on the bulk magnetic and transport
properties.
\end{abstract}


\maketitle

\section{Introduction}
The layered cobaltate Na$_{x}$CoO$_{2}$ has generated much recent
interest as a correlated electron material with unusual electronic
properties.  The structure of this compound consists of layers of
Co atoms, within an octahedral environment of oxygen atoms,
arranged on a triangular lattice.  The Na atoms form layers in
between the CoO$_2$ layers, and the Na stoichiometry can vary from
$x\sim 0.25$ to 0.85.  This family of materials has attracted much
recent interest due to the discovery of superconductivity below
4.5 K in hydrated Na$_{0.3}$CoO$_2$ $\cdot
1.3$H$_2$O.\cite{Takada03} The composition with $x \approx 0.75$
shows an unusually strong thermoelectric effect and has
Curie-Weiss magnetism coexisting with metallic
behavior.\cite{ray-prb99,chou-chi04} The electronic phase diagram
of non-hydrated Na$_x$CoO$_2$ is rich, with two metallic phases
existing for $x>0.5$ and $x<0.5$ which are separated by a
charge-ordered insulator at
$x=0.5$.\cite{Cava-PhaseDiagram03,Zandbergen04} It is clear that
the properties of Na$_x$CoO$_2$ can be strongly affected by the
mobility of the Na ions, which leads, for example, to the
formation of a Na superlattice at
$x=0.5$.\cite{Cava-PhaseDiagram03,Qing-04}  In order to further
understand the electronic and magnetic properties of
Na$_x$CoO$_2$, it is helpful to first understand the details of
the crystal structure.

In this paper we present neutron powder diffraction measurements
of samples of Na$_x$CoO$_2$ with $x$ near 0.75. We find that there
exist two distinct structures in this region of the phase diagram.
These structures differ from each other by the arrangement of the
Na ions. In addition, we find that the Na arrangement depends
sensitively on the temperature and doping level $x$. These results
are important because they demonstrate that samples of nominally
the same composition, $x\sim 0.75$, may have different physical
properties.  Recently, Sales and coworkers have found evidence for
a transition around 340 K using scanning calorimetry,
magnetization and transport measurements.\cite{Sales-Feb04}  They
attributed this behavior to a possible structural transition
involving Na ordering. Our powder neutron diffraction measurements
directly reveal how the crystal structure changes. We present
results from a systematic study involving neutron diffraction,
resistivity measurements, and susceptibility measurements on both
single crystal and powder samples annealed under various
conditions.

\section{EXPERIMENTAL}
Powder samples of Na$_x$CoO$_2$ were prepared using a solid-state
reaction technique.  A mixture of 0.75 Na$_{2}$CO$_{3}$ + 2/3
Co$_{3}$O$_{4}$ was used as the starting material.  The material
was reacted at elevated temperatures ($750^{\circ}$C and
$900^\circ$C) in air with repeated grindings over 12 hours cycles
until single phase Na$_{0.75}$CoO$_2$ was achieved, as verified
with x-ray diffraction.  The single crystal samples were grown via
the travelling solvent floating-zone method as described
previously.\cite{Chou2004} In this paper, we have examined the
structure of four different samples: (i) a crushed single crystal
labelled \textbf{CC}, (ii) an as-prepared powder which was
quenched in air from $900^\circ$ C labelled \textbf{P-900}, (iii)
an as-prepared powder quenched in air from $750^\circ$ C labelled
\textbf{P-750}, and (iv) a powder sample labelled
\textbf{P-anneal} which was annealed in a flowing oxygen
atmosphere while slowly cooled between from $900^\circ$ to
$50^{\circ}$C at a rate of $-1^{\circ}$C/min.  The samples'
histories are summarized in Table \ref{tab:table2}. All four
samples originated from different batches and therefore have
slight variations in stoichiometry $x$ near 0.75.

\squeezetable
\begin{table*}
\caption{\label{tab:table2}Summary of Na$_{0.75}$CoO$_2$ samples}
\begin{ruledtabular}
\begin{tabular}{ccccccccc}
name&batch\#&sample form&reaction atm.&highest reaction temp.&cooling\\
\hline
\textbf{CC}&$\# 2465$&crushed crystal&oxygen&$>1100^{\circ}$C&quenched\\
\textbf{P-900}&$\#2683$&powder&air&$900^{\circ}$C&quenched\\
\textbf{P-750}&$\#2683B$&powder&air&$750^{\circ}$C&quenched\\
\textbf{P-anneal}&$\#2683D$&powder&air+O$_{2}$ annealed&$900^{\circ}$C%
&slow cooled to $50^{\circ}$C\\
\end{tabular}
\end{ruledtabular}
\end{table*}

The neutron powder diffraction data of these samples were
collected using the BT-1 high-resolution powder diffractometer at
the NIST Center for Neutron Research, employing a Cu (311)
monochromator crystal to produce a beam of monochromatic neutrons
with wavelength $\lambda = 1.5403$~\AA.  Collimators with
horizontal angular divergences of $15^{\prime}, 20^{\prime}$, and
$7^{\prime}$ were used before and after the monochromator, and
after the sample, respectively. The scattered intensities were
measured as a function of 2$\theta$ in steps of 0.05$^{\circ}$ in
the range $3^{\circ}-168^{\circ}$. The structural parameters were
refined using the GSAS program.\cite{GSAS} The neutron scattering
lengths used in the refinements were 0.363, 0.253, and 0.581
($\times 10^{-12}$ cm) for Na, Co, and O, respectively.

Magnetization measurements were performed using a SQUID
magnetometer (Quantum Design MPMS-XL). The resistivity of a single
crystal sample was measured using the standard 4-probe technique.
Electrical contacts were attached to the surface of the sample
with silver epoxy and measurements were performed from 5~K to
350~K in applied fields of 0 Tesla and 14 Tesla using a Physical
Property Measurement System (Quantum Design).

\section{RESULTS}
\begin{figure}
\includegraphics[width=3.3in]{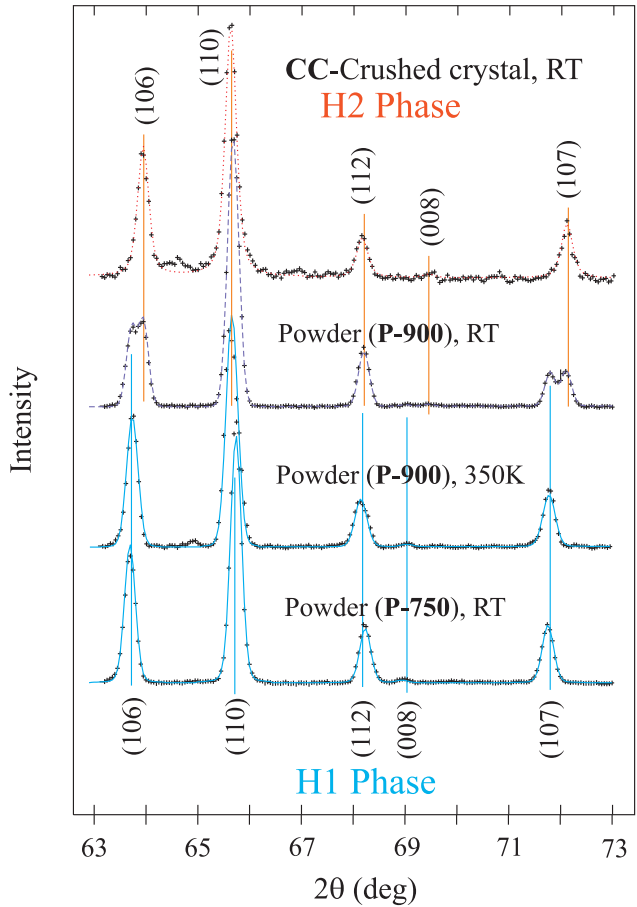}
\caption{\label{fig1} (color online) Portions of the neutron
powder diffraction pattern of Na$_x$CoO$_2$ samples annealed under
different conditions. The graphs from top to bottom correspond to
the following samples: crushed crystal \textbf{CC} measured at
room temperature (RT); powder \textbf{P-900} quenched in air from
$900^{\circ} C$ measured at RT; powder \textbf{P-900} measured at
350 K; powder \textbf{P-750} quenched in air from $750^{\circ}$C
measured at RT.  The curves are the calculated intensity with the
structural models H2 (red-dotted) and H1 (blue-solid),
corresponding to the crystal structures shown on Fig.
\ref{ModelBig}.}
\end{figure}

\begin{figure}
\includegraphics[width=3.1in]{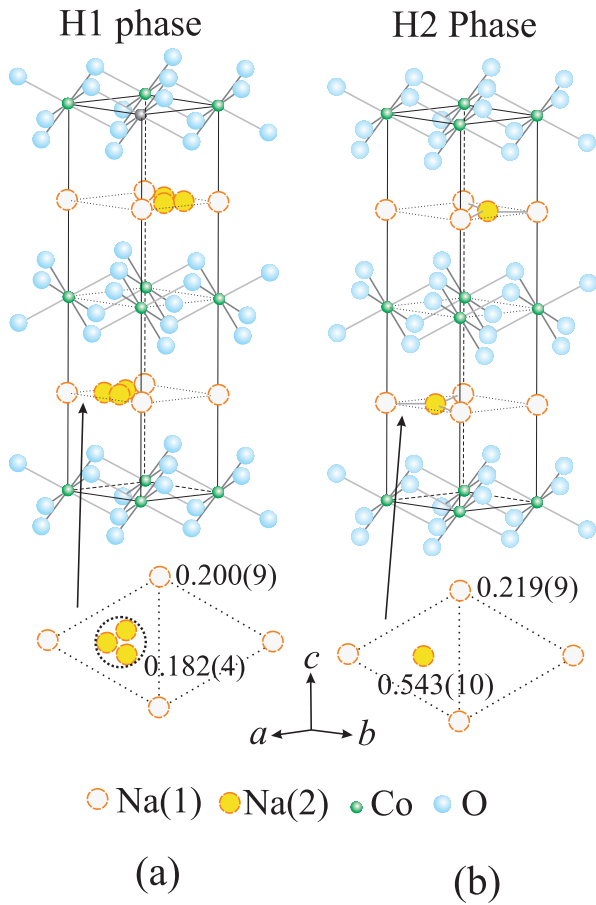}
\caption{\label{ModelBig} (color online) Models for the crystal
structure used in the refinements for Na$_x$CoO$_2$ ($x$ near
0.75), space group $P6_3/mmc$. The Na(2) atoms are (a) at the
$6h(2x, x, 1/4)$ site for structure H1 and (b) at the $2c(2/3,
1/3, 1/4)$ site for structure H2. The numbers shown in the lower
figure of the Na planes are the refined occupancy of each Na site
for the \textbf{P-900} sample.}
\end{figure}

We first present our results on the structural properties of our
samples which all have the same nominal Na concentration near
$x=0.75$. The structure of Na$_x$CoO$_2$ consists of layers of
edge-sharing CoO$_6$ octahedra, where the Co atoms form a
triangular lattice. For the hexagonal unit cell, the lattice
constants are $a_H \simeq 2.84$ and $c_H \simeq 10.82$ \AA.  The
Na ions occupy planes which lie in between the CoO$_2$ layers.
Previous measurements\cite{lynn:214516,jorgensen:214517} have
revealed two crystallographically distinct sites which the Na ions
occupy in the lattice, Na(1) and Na(2).  The relative population
of these sites varies from sample to sample and depends strongly
on the Na concentration $x$.

Figure 1 shows a portion of the neutron powder diffraction pattern
of Na$_{0.75}$CoO$_2$ for the crushed single crystal sample
(\textbf{CC}) and two of the powder samples (\textbf{P-900} and
\textbf{P-750}).  By comparing the room-temperature pattern for
sample \textbf{CC} (top-most profile) with the pattern for sample
\textbf{P-750} (bottom-most profile), we find that there are two
distinct structural phases.  As indicated in the plot, the
positions of several peaks (such as the (106) and (107) peaks) are
clearly different, indicating slight differences in the crystal
structure. The data can be fit using the hexagonal space group
$P6_3/mmc$ with structural models H1 or H2 as shown in Fig. 2.
\cite{lynn:214516,jorgensen:214517} These structural models will
be discussed in detail below.  The primary difference is that the
Na(2) ions reside at a higher symmetry position for structure H2
compared to structure H1.  The crushed crystal \textbf{CC} has
structure H2 at room temperature, whereas the powder sample
\textbf{P-750} has structure H1 at room temperature.

\begin{figure}
\includegraphics[width=3.3in]{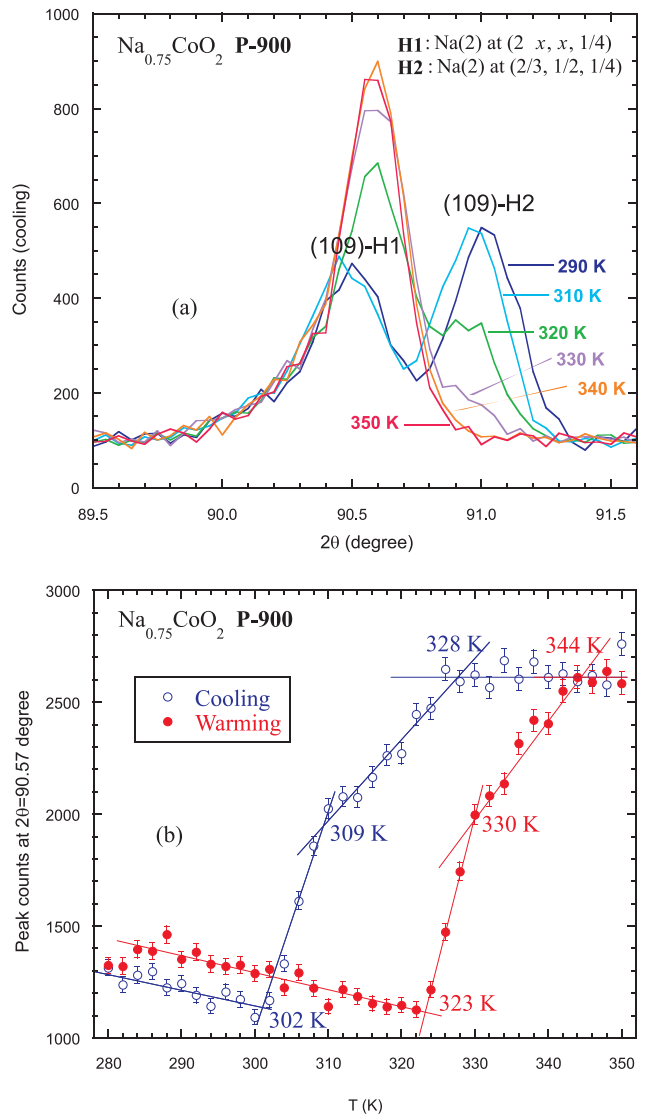}
\caption{\label{109}(color online) Top panel: scans through the
the (109) peak of sample \textbf{P-900} at various temperatures.
 Bottom panel: temperature dependence of the intensity of the
 (109)-H1 peak.  These data demonstrate that at high
temperatures (above ~340 K), model H1 is the stable structure. }
\end{figure}

The middle two diffraction profiles in Fig.~1 were taken on sample
\textbf{P-900} at room temperature and at $T=350$~K.
Interestingly, the room temperature pattern shows coexistence of
both structures, H1 and H2. When this sample is heated to 350~K,
the peaks associated with structure H2 disappear completely, and
the peaks associated with structure H1 are enhanced.  This
indicates that a structural phase transition occurs at an
intermediate temperature.  Note that the powder profile of the
\textbf{P-900} sample at 350 K is identical to the profile for the
\textbf{P-750} sample at room temperature, demonstrating that the
conversion of the phase with structure H2 to the phase with
structure H1 is complete. The diffraction pattern for the sample
\textbf{P-anneal}, which was slow-cooled in oxygen, shows that the
entire sample is described by structure H1. The structural
parameters for all four samples determined by Rietveld refinement
are summarized in Table~\ref{tab:table1}.

%%%%%%%%% Begin Table %%%%%%%%%%%%%%
\squeezetable
\begingroup
\begin{table*}
\caption{\label{tab:table1}Structural parameters for
Na$_x$CoO$_2$($x \simeq 0.75$). Space group $P6_3/mmc$. Atomic
positions: Co: 2a(0 0 0); Na(1): 2b(0 0 1/4); Na(2): 2c(2/3, 1/3,
1/4) for structure model A or 6h(2x, x, 1/4) for structure model
B; O: 4f(1/3, 2/3, z).\\}
\begin{ruledtabular}
\begin{tabular}{lllllllllll}
%\begin{tabular}{ccccccccccc}
Sample ID&&\multicolumn{3}{|c|}{\bf P-750}&\multicolumn{4}{c}{{\bf
P-900}  (~3.6wt\%CoO)}&\multicolumn{1}{|c|}{\bf P-anneal}
&\multicolumn{1}{c|}{\bf CC}\\
\cline{3-11}%\cline{5-11}
T (K)&&3.6&30&295&\multicolumn{2}{c|}{295}&\multicolumn{2}{c}{350}&295&295\\
\cline{6-9}
$x$(refined)&&0.721(8)&0.726(8)&0.711(8)&0.746(15)&\em{0.762(13)}%
&0.746&0.762&0.749(8)&\em{0.752(17)}\\
Structure&&{\bf H1}&{\bf H1}&{\bf H1}&{\bf H1}(46.2(2)\%)%
&{\bf H2}(\em{50.2(2)}\%)&{\bf H1}(42.9(7)\%)&{\bf H1}(53.4(6)\%)%
&{\bf H1}&{\bf H2}\\
a($\AA$)&&2.83607(4)&2.83603(4)&2.83709(4)&2.83709(5)&\em{2.84064(4)}%
&2.8369(1)&2.84250(9)&2.83628(5)&\em{2.84081(9)}\\
c($\AA$)&&10.8325(2)&10.8327(2)&10.8811(2)&10.8735(2)&\em{10.8117(2)}%
&10.8733(5)&10.8611(5)&10.8863(3)&\em{10.8115(5)}\\
V($\AA^3$)&&75.456(3)&75.455(3)&75.849(3)&75.796(3)&\em{75.554(3)}%
&75.785(5)&75.998(5)&75.842(4)&\em{75.562(6)}\\
Co&B($\AA^2$)&0.13(3)&0.14(2)&0.32(2)&0.20(3)&\em{0.20(3)}&0.20(3)%
&0.20(3)&0.38(4)&\em{0.13(7)}\\
Na(1)&B($\AA^2$)&0.31(8)&0.47(8)&1.04(9)&1.30(8)&\em{1.30(8)}&1.41(9)%
&1.41(9)&1.7(1)&\em{1.0(1)}\\
&$n$&0.196(5)&0.198(5)&0.198(5)&0.200(9)&\em{0.219(9)}&0.200&0.219%
&0.221(5)&\em{0.224(11)}\\
Na(2)&$x$&0.2810(7)&0.2814(7)&0.2834(8)&0.274(1)&&0.285(2)&0.299(2)%
&0.289(2)&\\
&B($\AA^2$)&0.31(8)&0.47(8)&1.04(9)&1.30(8)&\em{1.30(8)}&1.41(9)&1.41(9)%
&1.7(1)&\\
&n&0.175(2)&0.176(2)&0.171(2)&0.182(4)&\em{0.543(10)}&0.182&0.181&0.176(2)%
&\em{0.518(13)}\\
O&$z$&0.09054(7)&0.09052(6)&0.09023(7)&0.0904(1)&\em{0.0909(1)}&0.0925(2)%
&0.0888(2)&0.08990(8)&\em{0.0908(2)}\\
&B($\AA^2$)&0.43(2)&0.45(1)&0.61(1)&0.46(1)&\em{0.46(1)}&0.60(2)&0.60(2)%
&0.56(2)&\em{0.50(3)}\\
%\multicolumn{2}{c|}{295}&\multicolumn{2}{c}{350}&295&295
&$R_p$(\%)&4.75&4.62&4.42&\multicolumn{2}{c}{4.16}&\multicolumn{2}{c}{5.28}%
&4.90&\em{4.54}\\
&$R_{wp}$&5.90&5.78&5.39&\multicolumn{2}{c}{5.02}%
&\multicolumn{2}{c}{6.58}&6.54&\em{5.66}\\
&$\chi^2$&2.100&1.976&1.790&\multicolumn{2}{c}{1.608}%
&\multicolumn{2}{c}{2.149}&2.829&\em{1.323}\\
\end{tabular}
Selected bond ($\AA$) distances and angles (degree)\\
\begin{tabular}{lllllllllll}
Co-O&$\times6$&1.9087(4)&1.9085(4)&1.9097(4)&1.9103(6)&\em{1.9119(6)}%
&1.922(1)&1.904(1)&1.9077(5)&\em{1.912(1)}\\
O-Co-O&&95.97(3)&95.97(3)&95.94(2)&95.90(4)&\em{95.96(4)}&95.11(9)%
&96.59(8)&96.04(3)&\em{95.97(7)}\\
\\
Na(1)-O&$\times6$&2.3801(5)&2.3803(5)&2.3886(5)&2.3863(8)%
&\em{2.3768(8)}&2.370(2)&2.399(2)&2.3915(6)&\em{2.377(2)}\\
Na(2)-O&$\times4$&2.304(1)&2.305(1)&2.316(1)&2.302(2)&\em{2.3768(8)}%
&2.298(3)&2.346(4)&2.326(2)&\em{2.377(2)}\\
&$\times2$&2.564(3)&2.562(3)&2.563(3)&2.596(5)&\em{2.3768(8)}&2.540(8)%
&2.519(9)&2.547(5)&\em{2.377(2)}\\
Na(2)-O$_{average}$&&2.391(5)&2.389(5)&2.398(5)&2.400(8)&\em{2.3768(8)}%
&2.38(1)&2.40(2)&2.400(8)&\em{2.377(2)}\\
\end{tabular}
\end{ruledtabular}
\end{table*}
\endgroup
%%%%%%%%%%%% End Table %%%%%%%%%%%%%%%%%%%%
\begin{figure}
\includegraphics[width=3.5in]{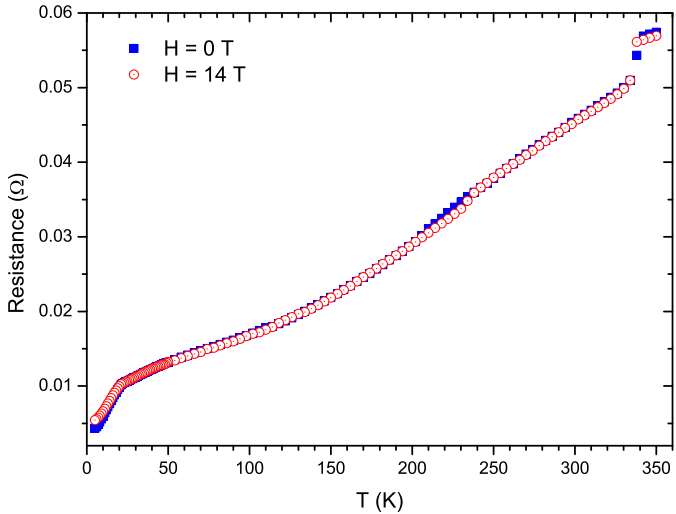}
\caption{\label{R} (color online) Resistivity versus temperature
on a single crystal prepared under the same conditions as the
crushed crystal sample \textbf{CC}.}
\end{figure}
\begin{figure}
\includegraphics[width=3.7in]{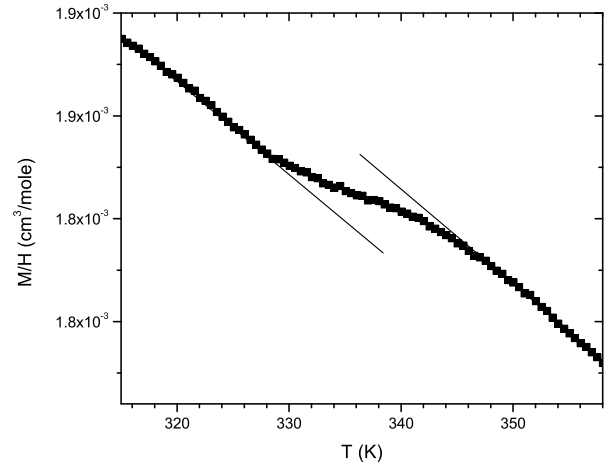}
\caption{\label{M} (color online) Magnetic susceptibility versus
temperature on the same sample as on Fig. \ref{R}. A magnetic
field of 5 T was applied approximately parallel to the
$ab$-plane.}
\end{figure}

In order to characterize the structural phase transition in more
detail, we measured the (109) Bragg peak of sample \textbf{P-900}
over a range of temperatures.  As shown in Figure~\ref{109}(a), at
a temperature of $T=290$~K there are two diffraction peaks,
(109)-H1 and (109)-H2, signifying the coexistence of both
structures (H1 and H2). As the temperature is raised in steps of
2~K, the intensity of the (109)-H2 peak decreases and the
intensity of (109)-H1 peak increases.  Upon warming, the sample
becomes single phase above 340 K, possessing only the H1
structure. Figure~\ref{109}(b) shows the intensity at a
$2\theta$-position corresponding to the (109)-H1 reflection as a
function of temperature. The intensity suddenly increases above
323~K upon warming, and reaches a maximum near 344~K.  Upon
cooling the intensity begins to decrease suddenly below 328~K and
levels off below 302~K.  The marked hysteresis (around 20 degrees
in temperature) suggests that the transformation from structure H2
to H1 is a first order transition.

The reason for the coexistence of the two structures at room
temperature is that sample \textbf{P-900} is likely composed of a
mixture of two stoichiometries with slightly different Na
contents.  The Rietveld refinement at $T=295$~K indicates the
sample is almost an equal mixture of two compounds with
$x=0.746(15)$ and $x=0.762(13)$.  (We note that the error bars for
$x$ for these two phases overlap.  Alternatively, we could label
these two phases using their $a$ and $c$ lattice constants, which
are clearly distinguishable.)  A complete diffraction pattern was
collected at 350 K and the data were analyzed by the Rietveld
method using a two-phase model with the same H1 structure but
different structural parameters. In this refinement the Na content
$x$ and the occupancies for each site were fixed at values
obtained from the 295 K data, assuming that the Na atoms do not
change sites (ie. Na(1) and Na(2)) between 295 and 350 K.  The
results are shown in Table~\ref{tab:table1}.  From these fits, we
can conclude that the temperature dependence plotted in
Fig.~\ref{109} is entirely derived from the phase associated with
$x=0.762$.

We have measured the transport properties of a single crystal
sample of Na$_{0.75}$CoO$_2$ which was prepared under the same
conditions as sample \textbf{CC}. Figure~\ref{R} shows the
resistance as a function of temperature between 5 K and 350 K,
measured in magnetic fields of 0 Tesla and 14 Tesla oriented along
the c-axis.  The most noticeable features of the resistivity
curves are the transitions near 22 K and 340 K.  Both of these
features agree with the reported measurements of Sales et.al. on a
floating-zone grown Na$_{0.75}$CoO$_2$ sample.\cite{Sales-Feb04}
The distinct jump in resistivity between 330~K and 340~K (upon
warming) in Fig.~\ref{R} closely matches the temperature range
(between 323~K and 344 K) over which the H2 structure becomes
transformed to the H1 structure for the sample shown in
Fig.~\ref{109}.  Since the crushed crystal is also described by
the H2 structure at room temperature, we conclude that the jump in
resistance near 340~K is a direct result of the structural
transition (H2$\rightarrow$H1) that we have discovered.  We also
observe a very small magnetoresistance effect between 200~K and
240~K.

Magnetic susceptibility measurements were performed on the same
single crystal used for the resistance measurement.  These data
are shown in Fig.~\ref{M}.  We find that there is a subtle change
in the temperature dependence between 330~K and 340~K.  This is
consistent with other reported measurements.\cite{Sales-Feb04}
Again, we can associate this behavior with the newly discovered
structural transition.  Measurements of the magnetic
susceptibility of the crushed crystal sample \textbf{CC} show
nearly identical behavior.

\section{DISCUSSION}
\begin{figure}
\includegraphics[width=3.3in]{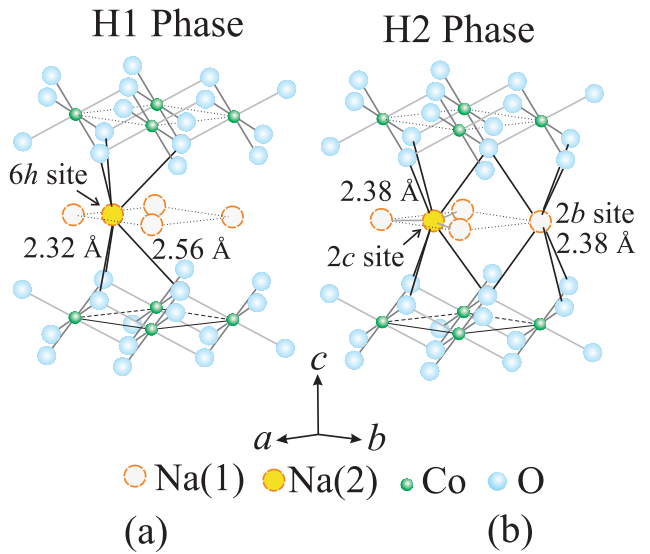}
\caption{\label{modelsmall}(color online) Coordination of Na(1)
and Na(2) in models H1 and H2.  The Na(1) ions are located at the
$2b$ site, center of NaO$_6$ triangular polyhedron, and are
coordinated to six oxygen atoms which are associated with two
nearest CoO$_6$ octahedra (directly above and below).  (a) The
Na(2) ions randomly occupy the 6h site, shifting away from the 2c
site, i.e. the center of the triangular polyhedron. (b)  The Na(2)
ions are located at the 2c site, the center of the triangular
mesh, and are coordinated to six oxygen atoms which are associated
with to six nearest CoO$_6$ octahedra (above and below CoO$_2$
blocks). }
\end{figure}

The most important finding of this work is the existence of two
distinct structures for Na$_x$CoO$_2$ samples with $x$ near 0.75.
The difference between the two structural models, H1 and H2, is
that in model H2 the Na(2) atoms occupy the $2c(2/3, 1/3, 1/4)$
site, while for the model H1 structure the Na(2) atoms randomly
occupy the $6h(2x, x, 1/4)$ site.  To emphasize the differences in
Na positions in the two models, we show a section of a Na layer,
projected along the c-axis, in the lower part of the
Fig.~\ref{ModelBig}. In addition, Fig.~\ref{modelsmall} shows the
coordination of the Na(1) and Na(2) sites. There are six oxygen
atoms surrounding the Na with the average Na-O bond distance $\sim
$2.38~\AA, forming a triangular polyhedron. The Na(1)O$_6$
triangular polyhedron is face sharing with two CoO$_6$ octahedra
(directly above and below) and the Na(2)O$_6$ triangular
polyhedron is corner sharing with six CoO$_6$ octahedra in the
nearest CoO$_2$ blocks.  Both Na(1) and Na(2) sites are partially
occupied and the ratio of the number of the Na ions is n$_{Na(2)}$
: n$_{Na(1)}$ = 2.5 : 1. No evidence of vacancy ordering was
observed in our experiments. In model H2, the Na(1) and Na(2) ions
are located at the corner and center of the triangular mesh of the
hexagonal lattice, respectively, and both Na(1) and Na(2) have
identical bonding distances to the nearest oxygen atoms.  In
contrast, in model H1, the Na(2) shifts away from the center of
the Co-ions triangle and randomly occupies the 6h(2x, x, 1/4)
sites. In this case, therefore, the Na ions shift away from the
center of the triangular NaO$_6$ polyhedron resulting in two long
(2.56~\AA) and four short (2.32~\AA) Na-O bonds.

\begin{figure}
\includegraphics[width=3.3in]{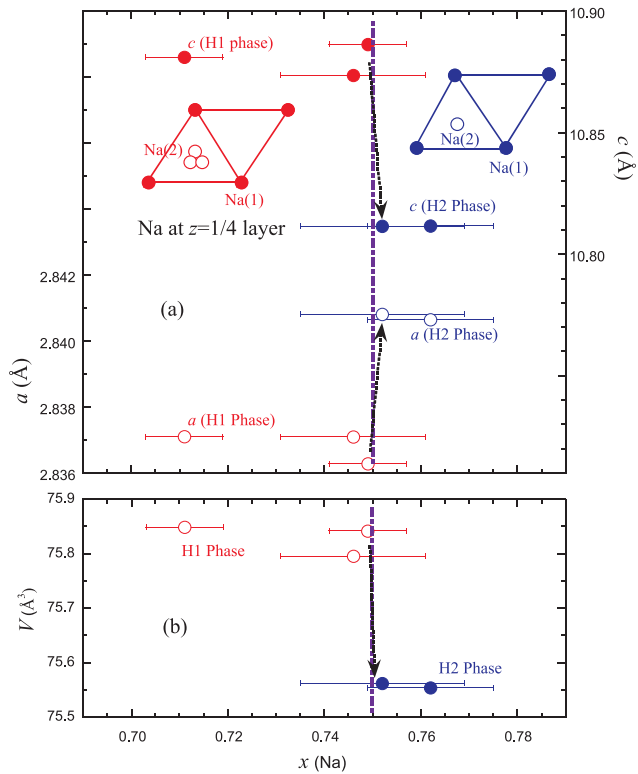}
\caption{\label{LatticeParam} (color online) Lattice parameters at
room temperature as a function of Na content $x$.  The change in
the lattice constants as $x$ increases above $x\simeq 0.75$
corresponds to the change from the H1 structure to the H2
structure.}
\end{figure}

Another important result from our work is shown in
Fig.~\ref{LatticeParam}. Here, we plot of the lattice constants as
a function of Na content $x$, where $x$ was determined by Rietveld
refinement as listed in Table \ref{tab:table1}. In samples with
structure H1, the a-axis is significantly shorter and the c-axis
is significantly longer than in samples with structure H2. Hence,
these two structures can be readily distinguished from each other.
Our refinements suggest that a sudden change occurs at $x$ near
0.75, in which structure H1 is the stable phase below
$x\simeq0.75$ and structure H2 is the stable phase above
$x\simeq0.75$. We note that the error bars for the refined values
of $x$ are large and overlap, except for the two samples:
\textbf{P-750} with $x=0.711$ and \textbf{P-900} with $x=0.762$.
However, the identification of $x=0.75$ as the critical
concentration is further confirmed by studies on other samples of
Na$_x$CoO$_2$ with different stoichiometry ($0.3 < x <
1$).\cite{Huang-unpubl} We note that the H1-to-H2 phase transition
is not observed in sample \textbf{P-750} which has $x=0.711$.  At
this value of $x$, the structure H1 is stable for the entire
temperature range measured.

The structural phase transition resembles an order-disorder phase
transitions in that Na atoms shift from a random occupation of the
6$h$ sites (structure H1) to a configuration in which every Na(2)
atom occupies a 2$c$ position (structure H2). The transition is
unusual in that it is the lower temperature phase (structure H2)
which has the higher symmetry for the Na(2) ions.  However, we
note that both structures have the same {\em average} space group
symmetry $P6_3/mmc$, as measured by neutron powder diffraction.

The sudden expansion of the $c$-axis lattice constant at the
transition from structure H1 to H2 may explain the sharp jump in
the resistance that we measured near 340~K upon warming. The
transport properties of Na$_x$CoO$_2$ are highly anisotropic, with
the resistivity within the $ab$-planes being several hundred times
higher than that perpendicular to the planes.\cite{Sales-Feb04}
The resistance of such an anisotropic material measured using the
4-probe technique usually contains an out-of-plane component due
to non-uniform current flow across the sample. Since the c-axis
resistivity should increase for a larger c-axis lattice constant,
this may explain the jump in our resistance data shown in
Fig.~\ref{R}.  Alternatively, the jump in resistivity may be
related to enhanced scattering due to the disorder in the H1
structure, which has randomly occupied $6h(2x, x, 1/4)$ sites.

In conclusion, we have found two distinct structural phases in
Na$_x$CoO$_2$ with $x$ near 0.75.  These two phase are
characterized by Na ions which occupy different sites,
$6h(2x,x,1/4)$ in structure H1 and $2c(2/3,1/3,1/4)$ in structure
H2.  The presence of either structure depends sensitively on the
specific conditions used during sample synthesis (such as
annealing environment and rate of cooling).  Our refinement
parameters indicate that this structural transition is especially
sensitive to the Na content $x$, with $x\simeq0.75$ as the
critical concentration separating the two phases. By raising the
temperature to around $T\simeq 323$~K, the high-symmetry structure
(H2) can be converted to the lower-symmetry structure (H1).  We
find that this structural transition also has signatures in
measurements of the bulk magnetic and transport properties.

\begin{acknowledgments}
The work at MIT was supported primarily by
the MRSEC Program of the National Science Foundation under grant
number DMR 02-13282. Identification of commercial equipment in the
text is not intended to imply recommendation or endorsement by the
National Institute of Standards and Technology.
\end{acknowledgments}

%%%%%%%%%%%%%%%%%% BIBLIOGRAPHY & END %%%%%%
\bibliography{NaCoO}

\begin{thebibliography}{12}
\expandafter\ifx\csname natexlab\endcsname\relax\def\natexlab#1{#1}\fi
\expandafter\ifx\csname bibnamefont\endcsname\relax
  \def\bibnamefont#1{#1}\fi
\expandafter\ifx\csname bibfnamefont\endcsname\relax
  \def\bibfnamefont#1{#1}\fi
\expandafter\ifx\csname citenamefont\endcsname\relax
  \def\citenamefont#1{#1}\fi
\expandafter\ifx\csname url\endcsname\relax
  \def\url#1{\texttt{#1}}\fi
\expandafter\ifx\csname urlprefix\endcsname\relax\def\urlprefix{URL }\fi
\providecommand{\bibinfo}[2]{#2}
\providecommand{\eprint}[2][]{\url{#2}}

\bibitem[{\citenamefont{Takada et~al.}(2003)\citenamefont{Takada, Sakurai,
  Takayama-Muromachi, Izumi, Dilanlan, and Sasaki}}]{Takada03}
\bibinfo{author}{\bibfnamefont{K.}~\bibnamefont{Takada}},
  \bibinfo{author}{\bibfnamefont{N.}~\bibnamefont{Sakurai}},
  \bibinfo{author}{\bibfnamefont{E.}~\bibnamefont{Takayama-Muromachi}},
  \bibinfo{author}{\bibfnamefont{F.}~\bibnamefont{Izumi}},
  \bibinfo{author}{\bibfnamefont{R.~A.} \bibnamefont{Dilanlan}},
  \bibnamefont{and} \bibinfo{author}{\bibfnamefont{T.}~\bibnamefont{Sasaki}},
  \bibinfo{journal}{Nature} \textbf{\bibinfo{volume}{422}}
  (\bibinfo{year}{2003}).

\bibitem[{\citenamefont{Ray et~al.}(1999)\citenamefont{Ray, Ghoshray, Ghoshray,
  and Nakamura}}]{ray-prb99}
\bibinfo{author}{\bibfnamefont{R.}~\bibnamefont{Ray}},
  \bibinfo{author}{\bibfnamefont{A.}~\bibnamefont{Ghoshray}},
  \bibinfo{author}{\bibfnamefont{K.}~\bibnamefont{Ghoshray}}, \bibnamefont{and}
  \bibinfo{author}{\bibfnamefont{S.}~\bibnamefont{Nakamura}},
  \bibinfo{journal}{Phys. Rev. B} \textbf{\bibinfo{volume}{59}},
  \bibinfo{pages}{9454} (\bibinfo{year}{1999}).

\bibitem[{\citenamefont{Chou et~al.}(2004{\natexlab{a}})\citenamefont{Chou,
  Cho, and Lee}}]{chou-chi04}
\bibinfo{author}{\bibfnamefont{F.~C.} \bibnamefont{Chou}},
  \bibinfo{author}{\bibfnamefont{J.~H.} \bibnamefont{Cho}}, \bibnamefont{and}
  \bibinfo{author}{\bibfnamefont{Y.~S.} \bibnamefont{Lee}}
  (\bibinfo{year}{2004}{\natexlab{a}}), \bibinfo{note}{cond-mat/0404061}.

\bibitem[{\citenamefont{Foo et~al.}(2003)\citenamefont{Foo, Wang, Watauchi,
  Zandbergen, He, Cava, and Ong}}]{Cava-PhaseDiagram03}
\bibinfo{author}{\bibfnamefont{M.~L.} \bibnamefont{Foo}},
  \bibinfo{author}{\bibfnamefont{Y.}~\bibnamefont{Wang}},
  \bibinfo{author}{\bibfnamefont{S.}~\bibnamefont{Watauchi}},
  \bibinfo{author}{\bibfnamefont{H.~W.} \bibnamefont{Zandbergen}},
  \bibinfo{author}{\bibfnamefont{T.}~\bibnamefont{He}},
  \bibinfo{author}{\bibfnamefont{R.~J.} \bibnamefont{Cava}}, \bibnamefont{and}
  \bibinfo{author}{\bibfnamefont{N.~P.} \bibnamefont{Ong}}
  (\bibinfo{year}{2003}), \bibinfo{note}{cond-mat/0312174}.

\bibitem[{\citenamefont{Zandbergen et~al.}(2004)\citenamefont{Zandbergen, Foo,
  Xu, Kumar, and Cava}}]{Zandbergen04}
\bibinfo{author}{\bibfnamefont{H.}~\bibnamefont{Zandbergen}},
  \bibinfo{author}{\bibfnamefont{M.}~\bibnamefont{Foo}},
  \bibinfo{author}{\bibfnamefont{Q.}~\bibnamefont{Xu}},
  \bibinfo{author}{\bibfnamefont{V.}~\bibnamefont{Kumar}}, \bibnamefont{and}
  \bibinfo{author}{\bibfnamefont{R.~J.} \bibnamefont{Cava}}
  (\bibinfo{year}{2004}), \bibinfo{note}{cond-mat/0403206}.

\bibitem[{\citenamefont{Huang et~al.}(2004{\natexlab{a}})\citenamefont{Huang,
  Foo, Lynn, Zandbergen, Lawes, Wang, Toby, Ramirez, Ong, and Cava}}]{Qing-04}
\bibinfo{author}{\bibfnamefont{Q.}~\bibnamefont{Huang}},
  \bibinfo{author}{\bibfnamefont{M.}~\bibnamefont{Foo}},
  \bibinfo{author}{\bibfnamefont{J.}~\bibnamefont{Lynn}},
  \bibinfo{author}{\bibfnamefont{H.}~\bibnamefont{Zandbergen}},
  \bibinfo{author}{\bibfnamefont{G.}~\bibnamefont{Lawes}},
  \bibinfo{author}{\bibfnamefont{Y.}~\bibnamefont{Wang}},
  \bibinfo{author}{\bibfnamefont{B.~H.} \bibnamefont{Toby}},
  \bibinfo{author}{\bibfnamefont{A.}~\bibnamefont{Ramirez}},
  \bibinfo{author}{\bibfnamefont{N.}~\bibnamefont{Ong}}, \bibnamefont{and}
  \bibinfo{author}{\bibfnamefont{R.}~\bibnamefont{Cava}}
  (\bibinfo{year}{2004}{\natexlab{a}}), \bibinfo{note}{cond-mat/0402255}.

\bibitem[{\citenamefont{Sales et~al.}(2004)\citenamefont{Sales, Jin, Affholter,
  Khalifah, Veith, and Mandrus}}]{Sales-Feb04}
\bibinfo{author}{\bibfnamefont{B.~C.} \bibnamefont{Sales}},
  \bibinfo{author}{\bibfnamefont{R.}~\bibnamefont{Jin}},
  \bibinfo{author}{\bibfnamefont{K.~A.} \bibnamefont{Affholter}},
  \bibinfo{author}{\bibfnamefont{P.}~\bibnamefont{Khalifah}},
  \bibinfo{author}{\bibfnamefont{G.~M.} \bibnamefont{Veith}}, \bibnamefont{and}
  \bibinfo{author}{\bibfnamefont{D.}~\bibnamefont{Mandrus}}
  (\bibinfo{year}{2004}), \bibinfo{note}{cond-mat/0402379}.

\bibitem[{\citenamefont{Chou et~al.}(2004{\natexlab{b}})\citenamefont{Chou,
  Cho, Lee, Abel, Matan, and Lee}}]{Chou2004}
\bibinfo{author}{\bibfnamefont{F.~C.} \bibnamefont{Chou}},
  \bibinfo{author}{\bibfnamefont{J.~H.} \bibnamefont{Cho}},
  \bibinfo{author}{\bibfnamefont{P.~A.} \bibnamefont{Lee}},
  \bibinfo{author}{\bibfnamefont{E.~T.} \bibnamefont{Abel}},
  \bibinfo{author}{\bibfnamefont{K.}~\bibnamefont{Matan}}, \bibnamefont{and}
  \bibinfo{author}{\bibfnamefont{Y.~S.} \bibnamefont{Lee}},
  \bibinfo{journal}{Phys. Rev. Lett.} \textbf{\bibinfo{volume}{92}}
  (\bibinfo{year}{2004}{\natexlab{b}}).

\bibitem[{\citenamefont{Larson and Von~Dreele}(1994)}]{GSAS}
\bibinfo{author}{\bibfnamefont{A.}~\bibnamefont{Larson}} \bibnamefont{and}
  \bibinfo{author}{\bibfnamefont{R.}~\bibnamefont{Von~Dreele}},
  \bibinfo{journal}{Los Alamos National Laboratory, Internal Report}
  (\bibinfo{year}{1994}).

\bibitem[{\citenamefont{Lynn et~al.}(2003)\citenamefont{Lynn, Huang, Brown,
  Miller, Foo, Schaak, Jones, Mackey, and Cava}}]{lynn:214516}
\bibinfo{author}{\bibfnamefont{J.~W.} \bibnamefont{Lynn}},
  \bibinfo{author}{\bibfnamefont{Q.}~\bibnamefont{Huang}},
  \bibinfo{author}{\bibfnamefont{C.~M.} \bibnamefont{Brown}},
  \bibinfo{author}{\bibfnamefont{V.~L.} \bibnamefont{Miller}},
  \bibinfo{author}{\bibfnamefont{M.~L.} \bibnamefont{Foo}},
  \bibinfo{author}{\bibfnamefont{R.~E.} \bibnamefont{Schaak}},
  \bibinfo{author}{\bibfnamefont{C.~Y.} \bibnamefont{Jones}},
  \bibinfo{author}{\bibfnamefont{E.~A.} \bibnamefont{Mackey}},
  \bibnamefont{and} \bibinfo{author}{\bibfnamefont{R.~J.} \bibnamefont{Cava}},
  \bibinfo{journal}{Physical Review B} \textbf{\bibinfo{volume}{68}}
  (\bibinfo{year}{2003}).

\bibitem[{\citenamefont{Jorgensen et~al.}(2003)\citenamefont{Jorgensen, Avdeev,
  Hinks, Burley, and Short}}]{jorgensen:214517}
\bibinfo{author}{\bibfnamefont{J.~D.} \bibnamefont{Jorgensen}},
  \bibinfo{author}{\bibfnamefont{M.}~\bibnamefont{Avdeev}},
  \bibinfo{author}{\bibfnamefont{D.~G.} \bibnamefont{Hinks}},
  \bibinfo{author}{\bibfnamefont{J.~C.} \bibnamefont{Burley}},
  \bibnamefont{and} \bibinfo{author}{\bibfnamefont{S.}~\bibnamefont{Short}},
  \bibinfo{journal}{Physical Review B} \textbf{\bibinfo{volume}{68}}
  (\bibinfo{year}{2003}).

\bibitem[{\citenamefont{Huang et~al.}(2004{\natexlab{b}})\citenamefont{Huang,
  Foo, Lynn, Toby, {R. A. Pascal Jr.}, Zandbergen, and Cava}}]{Huang-unpubl}
\bibinfo{author}{\bibfnamefont{Q.}~\bibnamefont{Huang}},
  \bibinfo{author}{\bibfnamefont{M.}~\bibnamefont{Foo}},
  \bibinfo{author}{\bibfnamefont{J.}~\bibnamefont{Lynn}},
  \bibinfo{author}{\bibfnamefont{B.}~\bibnamefont{Toby}},
  \bibinfo{author}{\bibnamefont{{R. A. Pascal Jr.}}},
  \bibinfo{author}{\bibfnamefont{H.}~\bibnamefont{Zandbergen}},
  \bibnamefont{and} \bibinfo{author}{\bibfnamefont{R.}~\bibnamefont{Cava}}
  (\bibinfo{year}{2004}{\natexlab{b}}), \bibinfo{note}{unpublished}.

\end{thebibliography}

\end{document}